\documentclass[preprint,prl]{revtex4}

\usepackage{graphicx}
\usepackage{dcolumn}
\usepackage{bm}

\begin{document}

\setlength{\unitlength}{1mm}
\textwidth 15.0 true cm
\textheight 22.0 true cm
\headheight 0 cm
\headsep 0 cm
\topmargin 0.4 true in
\oddsidemargin 0.25 true in

\title{Three Form Inflation}

\author{Andrei Gruzinov}
 \affiliation{Center for Cosmology and Particle Physics, Department of Physics, New York University, NY 10003}

\date{January 25, 2004}

\begin{abstract}

It is noted that inflation driven by a three form with arbitrary self coupling satisfies a consistency relation $n_S-1=2n_T$, between spectral indices of scalar and tensor cosmological perturbations. The standard consistency relation for the ratio of the amplitudes of perturbations is only valid for quadratic self coupling of the three form.

\end{abstract}


\maketitle

\section{Introduction}

It is not known what are the fields that drive inflation. The simplest models postulate that inflation is driven by a self-coupled scalar field \cite{standard1,standard2}. The only way to ascertain that inflation was indeed driven by a scalar is through a consistency relation, $P_T/P_S=-8n_T$, where $P_{T,S}$ are the spectra of tensor and scalar perturbations, $n_T$ is the spectral index of the tensor perturbations. Until this consistency relation is confirmed, one should keep an open mind. It has been suggested that inflation can be driven by a scalar field with non-minimal kinetic term \cite{k1}, or by a self-coupled vector field \cite{v}. 

We note here that inflation can be driven by a self-coupled three form. The three form inflation satisfies a different consistency relation, $n_S-1=2n_T$, where $n_S$ is the spectral index of the scalar perturbations. At the same time, the standard consistency relation is valid only if the self-coupling of the three form is quadratic.

\section{The Three Form}
Let $A^{\mu}$ be the dual of the three form. The action is
\begin{equation}
S={1\over 2}\int d^4x \sqrt{-g}\left( (\nabla _{\mu} A^{\mu})^2-F(A_{\mu}A^{\mu})\right),
\end{equation}
where $\nabla _{\mu}$ is covariant derivative, and $F(A^2)$ is an arbitrary self-coupling.
The field satisfies the equation of motion
\begin{equation}
\partial _\mu \nabla \cdot A~+~F'A_\mu=0,
\end{equation}
where $F'\equiv dF/dA^2$.
The energy-momentum of the three form is then
\begin{equation}
T^{\mu \nu}=F'A^\mu A^\nu -{1\over2}g^{\mu \nu}\left( (2F'A^2-F)-(\nabla \cdot A)^2\right).
\end{equation}

It follows from the equation of motion that a scalar $\phi \equiv -\nabla \cdot A$ satisfies 
\begin{equation}
\nabla \cdot \left( {1\over F'}\nabla \phi \right) +\phi =0,
\end{equation}
with the argument of $F'$ given implicitly by 
\begin{equation}\label{Aphi}
F'^2A^2=(\partial \phi )^2. 
\end{equation}
In terms of $\phi$, the energy-momentum is
\begin{equation}
T^{\mu \nu}={1\over F'} \partial ^\mu \phi \partial ^\nu \phi -{1\over2}g^{\mu \nu}\left( (2F'A^2-F)-\phi ^2 \right).
\end{equation}
This energy-momentum and the $\phi$ equation of motion follow from the action 
\begin{equation}
S={1\over 2}\int d^4x \sqrt{-g}{\cal L} ,
\end{equation}
with
\begin{equation}\label{lagr}
{\cal L} ={1\over 2}\left( (2F'A^2-F)-\phi ^2\right) .
\end{equation}
Thus, according to eqs. (\ref{lagr},\ref{Aphi}), the self-coupled three form is equivalent to a scalar with non-minimal kinetic term and quadratic potential. For linear $F$, eqs. (\ref{lagr},\ref{Aphi}) give the Lagrangian of a massive scalar field. This proves that there do exist self-couplings of the three form that provide an observationally acceptable, \cite{teg}, inflation.

\section{The Three Form Inflation}
Inflation driven by a scalar with arbitrary local Lagrangian density ${\cal L}= {\cal L}(\partial \phi ,\phi )$ was considered in \cite{k1,k2}. All we need to do is to apply their results to a special case 
\begin{equation}
{\cal L} =f(X)-\phi ^2/2,
\end{equation}
where $f$ is an arbitrary function of $X\equiv (\partial \phi)^2/2$. We will follow the notations of \cite{k2}.

For spatially flat FRW, the Friedmann equations are
\begin{equation}
H^2\propto \epsilon ,
\end{equation}
\begin{equation}
\dot{ \epsilon }=-3H(\epsilon +p)=-6HXf',
\end{equation}
where $X\equiv \dot{\phi }^2/2$, and $'\equiv d/dX$. For slow roll, $\epsilon\approx \phi ^2/2$, and it follows that slow roll inflation occurs at a constant value of $X$. 

Since the speed of sound $c_s$ and $\epsilon +p$ depend only on $X$, equation (35) of \cite{k2} gives the scalar index
\begin{equation}
n_S-1=-6\left( 1+{p\over \epsilon }\right).
\end{equation}
At the same time (equation (38) of \cite{k2})
\begin{equation}
n_T=-3\left( 1+{p\over \epsilon }\right),
\end{equation}
giving the consistency relation 
\begin{equation}
n_S-1=2n_T.
\end{equation}

\begin{acknowledgments}
I thank Gia Dvali and Gregory Gabadadze for useful discussions. This work was supported by the David and Lucile Packard Foundation.
\end{acknowledgments}

\end{document}